\documentclass[conference]{IEEEtran}
\IEEEoverridecommandlockouts
\renewcommand{\thesubsection}{\thesection.\alph{subsection}}
\usepackage{cite}
\usepackage{amsmath,amssymb,amsfonts}
\usepackage{algorithmic}
\usepackage{graphicx}
\usepackage{textcomp}
\usepackage{xcolor}
\usepackage{enumitem}
\usepackage{float}
\usepackage{booktabs}
\usepackage{graphicx}
\usepackage{fancyhdr}
\usepackage{subcaption}
\usepackage{caption}
\usepackage[multiple]{footmisc}
\usepackage{hyperref}
\hypersetup{
    colorlinks=true,
    linkcolor=blue,
    filecolor=magenta,      
    urlcolor=blue,
    pdftitle={Sharelatex Example},
    bookmarks=true,
    pdfpagemode=FullScreen,
}
\let\oldbibliography\thebibliography
\renewcommand{\thebibliography}[1]{%
  \oldbibliography{#1}%
  \setlength{\itemsep}{0pt}%
}

\ifCLASSINFOpdf
\else
\fi
\begin{document}
%
\title{A Survey of Multimedia Technologies and Robust Algorithms}

\author{\IEEEauthorblockN{Zijian Kuang}
\IEEEauthorblockA{Department of Computing Science\\
University of Alberta\\
Edmonton, Canada \\
Email: kuang@ualberta.ca}
\and
\IEEEauthorblockN{Xinran Tie}
\IEEEauthorblockA{Department of Computing Science\\
University of Alberta\\
Edmonton, Canada \\
Email: xtie@ualberta.ca}
}


%


\maketitle

\begin{abstract}
Multimedia technologies are now more practical and deployable in real life, and the algorithms are widely used in various researching areas such as deep learning, signal processing, haptics, computer vision, robotics, and medical multimedia processing \cite{multimedia_survey}. This survey provides an overview of multimedia technologies and robust algorithms in multimedia data processing, medical multimedia processing, human facial expression tracking and pose recognition, and multimedia in education and training. This survey will also analyze and propose a future research direction based on the overview of current robust algorithms and multimedia technologies. We want to thank the research and previous work done by the Multimedia Research Centre (MRC), the University of Alberta, which is the inspiration and starting point for future research. 
\end{abstract}


%
\IEEEpeerreviewmaketitle

\section{Introduction}
Today, multimedia technologies have been widely studied in various researches and have been applied to people’s life intelligently. As a result, within the fields of human facial research, human action study, medical research, and data processing research, different multimedia algorithms have been proposed to utilize current resources to enhance multimedia usages and performance further. 

In this survey, we present a discussion on how the series of MRC’s multimedia algorithms and solutions can be used to make future research and our lives easier and safer. The idea of this survey also works on promoting more exchanges between researchers from various communities and helping initiate new interdisciplinary collaborations. Ultimately, MRC covers a broad range of disciplines in multimedia technologies and is driven to accelerate more projects that need expertise in multiple disciplines.

\section{Multimedia data processing}
Multimedia data, in general, can include images, videos, audio, and motion capture data. The data are widely used in image fusion, stereo matching, data compression, and videoconferencing applications. The multimedia data processing based algorithms proposed by MRC include:
\subsection{QoE-Based Multi-Exposure Fusion in Hierarchical Multivariate Gaussian CRF}
In 2013, A. Basu, I. Cheng, and R. Shen proposed a Hierarchical Multivariate Gaussian Conditional Random Field (HMGCRF) model, which was based on perceptual quality measures to provide viewers better quality experience on fused images. Since many state-of-art fusion methods have not considered the impact of perceptual factors while designing the image processing algorithm, HMGCRF takes account of the human visual system and uses perceived local contrast and color saturation to improve the performance of Multi-Exposure Fusion (MEF). 

The proposed model, HMGCRF, applies a novel MEF method that exploits both contrast and color information to deliver maximum image details. Using the MEF techniques instead of HDR imaging techniques, minimal user intervention is required, and a visually appealing and improved image is directly built. The human visual system's probability is brought into the HMGCRF model to deliver maximum image details. With a source image is given as input, the individual pixel’s contribution to the fused image is perceptually tuned by perceived local contrast and color saturation. Perceived local contrast that works in the LHS color space's luminance channel is used to make sure well-exposed regions contribute more to the fused image.

In contrast, saturation definition is used to measure the colorfulness of a pixel. The first time modeling the probability for human eyes is brought into multi-exposure fusion to detect local contrast. The maximum local detail perseveration is achieved using HMGCRG with perceptual quality measures while exhibiting more vivid colors on fused images simultaneously \cite{HMGCRF}.

\subsection{Stereo Matching Using Random Walks}
In 2008, R. Shen, I. Cheng, X. Li, and A. Basu presented a novel random walks framework based stereo matching algorithm. The model first extracts a set of reliable matching pixels. The matching is based on the prior matrices and Laplacian matrices on the information of adjacent pixels. Then the model uses the set of pixels as seeds to determine the disparities of unreliable regions by solving a Dirichlet problem. When building the Laplacian matrices and prior matrices, the lightning and illumination variance is also considered to improve the resulting disparity maps' accuracy \cite{stereo_matching_random_walk}.

The algorithm proposed in the paper first estimate the disparity for both images by solving the equation below \cite{stereo_matching_random_walk}:

\begin{equation}
\left(L+\gamma \sum_{r=1}^{k} \Lambda^{r}\right) x^{s}=\lambda^{s}
\end{equation}

where $\boldsymbol{\Lambda}=\operatorname{diag}\left(\boldsymbol{\lambda}^{\boldsymbol{s}}\right)$ \cite{stereo_matching_random_walk}.

Then a left-right checking is applied to determine a set of reliable matching pixels by \cite{stereo_matching_random_walk}:

\begin{equation}
\left(p_{l}, d_{l}\right) \in\left\{\begin{array}{ll}
\boldsymbol{M}, & d_{l} \in\left[-d_{r}-\theta_{3},-d_{r}\right] \\
\boldsymbol{U}, & \text { otherwise }
\end{array}\right.
\end{equation}

where $\theta_{3}=\left\lceil\frac{d_{\max }^{\prime}}{7}\right\rceil$ presents the maximum disparity in the initial disparity maps. Since the pixels occluded in the second image Ir, they will be labeled as unreliable without the left-right checking \cite{stereo_matching_random_walk}.

After left-right checking, the algorithm will compute the optimal disparity map by using the intermediate result. The algorithm also uses $Y C_{b} C_{r}$ space to automatically update the parameters to increase image matching accuracy due to the illumination condition variance \cite{stereo_matching_random_walk}.

\subsection{Perceptually Guided Fast Compression of 3D Motion Capture Data}
Motion capture data is widely used in producing skeletal animations, and efficient compression of motion capture data is important in optimizing the environment's usage with a limited amount of bandwidth and memory and preserving the high quality of animation. In 2001, A. Firouzmanesh, I. Cheng, and A. Basu proposed a method to perform a better compression ratio with shorter compression and decompression time. The perceptually guided fast compression model is proposed to optimize the coefficient selection algorithm considering two critical factors: the bone's length connected to a joint and the variation in rotation \cite{perceptually_guided}. This paper also inspires the other researchers that, other than the two factors like bone length and variation in rotation, there can be many other factors affecting the quality of animations, such as the distance from camera, horizontal and vertical velocity of the object, and the size of the limbs \cite{perceptually_guided}.

\subsection{Variable Resolution Teleconferencing}
With the rapid development of computers, a novel image compression scheme based on Variable Resolution (VR) sensing is proposed. In the paper, \cite{vr_teleconf1}, A. Basu et al. utilized VR in the area of videoconferencing. VR guarantees the compression ratio, can be performed efficiently, and allows further compression using other methods. For each image, there is an area that contains greater interest and is named fovea. Under the VR transform, a pixel is moved certain units away from the fovea. This transformation is easily reversed and ensures that points that are at the maximum distance from the fovea in the original image also have the maximum distance in the VR image. However, the images transformed by the basic VR model are not in the shape of rectangular. Multiple scaling factors are used to map the transformed image to a rectangle and utilize the full space to solve this problem. Also, multiple foveas cases are solved by calculating the locations for points using each fovea separately. The true location is later confirmed with the weighting function based on the distance from the original point to the fovea. For interpolation, VR has used straight pixel sampling, which selects one pixel to represent its nearest neighbors. In conclusion, the advantages of variable resolution enable the high quality of teleconferencing for machines with limited speed, and the frame rates are maintained without additional hardware.

\subsection{Videoconferencing using Spatially Varying Sensing with Multiple and Moving Foveae}
In the paper \cite{vr_teleconf2}, A. Basu et al. proposed a new method for videoconferencing using spatially varying sensing. Variable Resolution (VR) is also introduced to combine information from multiple points of interest (foveae) in one image. By identifying the fovea, it can be used to tracking a moving object of interest. VR is newly applied to image compression and is adept at controlling samples, guaranteeing compression ratios, performing efficiently, and working with other methods. Under the VR transform, a pixel with polar pixel (r, $\theta$) is transformed to (vr, $\theta$) where 
\begin{equation}
v r=\ln (r * \alpha+1) * s f
\end{equation}
Since pixels are discrete, one VR pixel may represent several pixels after the transformation. However, the resulted image after compression is non-rectangular. To solve this issue, a new approach is introduced to simplify the formula significantly by isolating the vertical and horizontal components with lower computational complexity \cite{vr_teleconf2}. For moving fovea in a sequence of images, the fovea's location is tracked using a look-up table. The look-up table structure is optimized to contain only the result of transform for the points in the first quadrant. Moreover, both cooperative and competitive foveae are examined to define multiple foveae. In conclusion, VR is ideal for teleconferencing due to its power of compressing images, maintaining framerates, and saving hardware cost. 

\subsection{Modeling Fish-eye Lenses}
To overcome expensive computation and the requirement of special sensors from existing variable-resolution transforms, A. Basu et al. proposed two fish-eye transformations (FET) in the paper \cite{fish-eye}. The proposed transforms are based on fish-eye lenses' characteristics and easy to implement by using off-the-shelf lenses. Since the original and complex logarithmic mapping can not maintain the image continuity across the vertical meridian and not be recovered easily, a simplifies VR projection is introduced in the paper. By simplifying the projection, the image continuity is preserved, but it also produces anisotropic distortions in peripheral regions \cite{fish-eye}. However, this approximation can be easily implemented by utilizing a wide-angle or fish-eye lens. Polynomial fish-eye transform (PFET) is another proposed method. It models the data by adjusting the parameters to describe the actions of condensing and summarizing a set of data. The difference in FERT compared with FET is the polynomial function is used instead for the transformation. In conclusion, simple transformations are introduced to model VR images. The proposed model is easy to implement using a fish-eye lens and presented to be closely fitted with a real fish-eye lens.

\section{Medical multimedia processing}
Medical multimedia information includes conventional chest radiograph (CXR), CT, and MRI. Medical multimedia information processing aims to perform medical image segmentation and analysis in a fully automatic way to provide more accurate medical diagnostics information. The medical multimedia processing related papers proposed by MRC include:

\subsection{Gradient Vector Flow based Active Shape Model for Lung Field Segmentation in Chest Radiographs} 
In 2009, T, Xu, M. Mandal, R. Long, and A. Basu proposed a modified gradient vector flow-based active shape model (GVF-ASM) using a global nonlinear function GVF field for lung field segmentation in chest radiographs \cite{gradient_segmentation}. 

To improve the accuracy and reduce the complexity of segmentation, the authors proposed a new points evolution equation as follow \cite{gradient_segmentation}:

\begin{equation}
C=C^{\prime}+w \cdot \operatorname{sgn}\left(g\left(C^{\prime}\right)\right) \cdot e^{-\mid g\left(c^{\prime}\right)}
\end{equation}

Sign function sgn is used to ensure the gradient vector’s direction and the exponential function exp(-|g(C’)|) works as a smooth monotonically decreasing function, including the points for those strong edges. After the finding, all the control points, the shape model’s parameters can be updated with a certain constraint. The stopping point could be iterations numbers or a threshold of the points’ Euclidean distance between two consecutive iterations \cite{gradient_segmentation}.

\subsection{Airway Segmentation and Measurement in CT Images}
In this literature, A. Basu, I. Cheng et al. introduced a methodology for airways segmentation by using Cone Beam CT data and measuring airways' volume. Since many of the researches focus on the segmentation and reconstruction of lower airway trees, a new strategy for detecting the boundary of slices of the upper airway and tracking the airway's contour using Gradient Vector Flow (GVF) snakes is introduced. 

The traditional GVF snakes are not performing well for airway segmentation when applied directly to CT images. Therefore, this paper's new method has modified the GVF algorithm with edge detection and sneak-shifting steps. By applying edge detection before the GVF snakes and using snake shifting techniques, the prior knowledge of airway CT slices is utilized, and the model works more robustly. The previous knowledge of the shape of the airway can automatically detect the airway in the first slice. The detected contour will then be used as the second slice's sneak initialization and so on \cite{airway}. A heuristic is also applied to differentiate bones from the airway by the color to make sure the snake converges correctly. Following this, the airway volume is estimated based on the 3D model constructed with automatically detected contours.

\subsection{Automatic Segmentation of Spinal Cord MRI Using Symmetric Boundary Tracing} 
In 2010, D. P. Mukherjee, I. Cheng, N. Ray, V. Mushahwar, M. Lebel, and A. Basu proposed an adaptive active contour tracing algorithm to extract spinal cord from MRI in a fully automatic manner. The proposed segmentation method first draws a circle at the center of the spinal cord region. Since from anatomy, the axis of symmetry of the spinal is a unique diameter that can divide the circle into two reflective symmetry to the half circles and can be further divided into N sectors. The proposed model then uses the Bhattacharya coefficient (BC) between the intensity histograms of the two sectors as below \cite{automatic_segmentation}:

\begin{equation}
\mathrm{Score}\left(C_{1}, C_{2}\right)=\sum_{i=1}^{N} \mathrm{BC}\left(h_{i}^{1}, h_{i}^{2}\right)
\end{equation}

The $h_{i}^{1}$ and $h_{i}^{2}$ indicate the intensity histograms of two corresponding sectors $C_{1}$ and $C_{2}$. Then the highest core value can be considered as the symmetry axis of the circle. The alignment and unskewing process can be further performed based on the symmetry axis with the vertical y-axis. Then the active tracing algorithm is used to segment the muscle region as below \cite{automatic_segmentation}:

\begin{equation}
c(p, q)=\exp \left(\frac{-\left(|\nabla I|_{p}+|\nabla I|_{q}\right)^{2}}{2 S_{m}^{2}}+\frac{\left(\left|\nabla I^{\theta}\right|_{p}-\left|\nabla I^{\theta}\right|_{q}\right)^{2}}{2}\right)
\end{equation} 

The $\nabla I$ indicates the gradient magnitude, and $\nabla I^{\theta}$ indicates the gradient orientation of the image I. The active tracing uses minimum cost path c(p, q) to perform a directed graph search for the optimal path in the energy surface defined by c(p, q) \cite{automatic_segmentation}.

The authors further elaborate on the active tracing to implement a unique contour energy minimization surface to increase the accuracy \cite{automatic_segmentation}.

\section{Human facial expression tracking and pose recognition}
Human facial expression tracking includes eye-tracking and nose shape estimation and can be detected with an active camera. Human pose estimation includes the stages of media skeletonization, spatial mapping and poses decision. The human facial expression tracking and pose recognition related papers proposed by MRC include:

\subsection{Eye Tracking and Animation for MPEG-4 Coding}
A. Basu et al. proposed a heuristic that can effectively improve facial feature detection, track algorithm, and focus on eye movements. Since accurate localization and tracking of facial features are essential to high-quality model-based coding (MPEG-4) systems, it is essential to automatically detect and track facial features and synthesis and facial expression analysis. In this paper, an improvement in the initial localization process and simple processing of images with Hough transform and deformable templates are used to produce more accurate results \cite{eye_tracking}. By introducing exploitation of the color information of eyes, the feature detection becomes more robust and faithful.  Moreover, a methodology for eye movement synthesis is also presented in this paper. After extracting iris and eyelids' contours from the image sequence, the deformation of the 3D model's eyes is computed and used to synthesize the real eye expression completely.  Further extension of the approach will be applied to lip movements and network strategies for a real-time system \cite{eye_tracking}.

\subsection{Nose Shape Estimation and Tracking for Model-Based Coding}
In previous researches, most facial feature extraction methods are focused on eye and mouth feature extractions. In contrast, the detection of the nostril and nose-side shape can be used in facial expression recognition since the different expressions can result in the change of nose shape. In 2001, A. Basu and L. Yin proposed a feature detection method that focused on the nose shape recognition \cite{nose_shape}. First, a two-stage region growing algorithm is developed to extract the feature blobs. Then two deformable templates are pre-defined to detect and track the nostril shape and nose-side shape. In the last step, a 3D wireframe model is matched onto the individual face to track the facial expressions using energy minimization \cite{nose_shape}.

\subsection{Integrating Active Face Tracking with Model-Based Coding}
In 1999, A. Basu and L. Yin proposed a system to detect and track a talking face with an active camera and then implement adaptation and animation automatically onto the detected face \cite{face_tracking}. This paper proposes an advanced head silhouette generation method to detect the motion of a talking face. A spatiotemporal filter is used to fuse the motion mask and complete the moving head detection. After the head region is detected, the deformable template matching combined with color information extraction and Hough Transform is used to extract the facial features such as eyes and mouth. In the end, a 3D wireframe model is generated to fit onto the moving face so that the positions of the eyes and mouth can be determined along with the entire face. They further introduced a “coarse-to-fine” adaptation algorithm to complete the face features adaptation process. \cite{face_tracking}.

\subsection{Pose Recognition using the Radon Transform}
In 2005, M. Singh, M. Mandal, and A. Basu proposed a novel method for human pose recognition using the Radon Transform. The proposed algorithms are based on two assumptions: first, any pose of the legs or arms can be replaced by the medial skeleton representation of the pose; second, the representation of a pose in the Rado Transform space includes the information regarding the pose classification \cite{pose_rt}.

To perform pose estimation, the proposed model includes five stages: image acquisition and preprocessing, media skeletonization, Radon Transform computation, spatial maxima mapping (SMM) algorithm, and pose decision \cite{pose_rt}.

In image acquisition and preprocessing, a background separation using a statistical background modeling approach is performed to generates binary images of human poses. Next, the medial skeletonization is utilized with a special discrete function that uses 1 to indicate if the part is skeleton and 0 for others. After skeletonization, the Radon Transform is performed to detect the orientation of the lines using the equation below \cite{pose_rt}:

\begin{equation}
\Re_{f}[r, \alpha]=\sum_{x=0}^{X-1} \sum_{j=0}^{1 Y-1} f_{s}(x, y) \delta(r-x \cos \alpha-y \sin \alpha)
\end{equation}

where $\delta(.)$ denots the Kronecker delta function. Thresholding and dilation are also applied to extract the most significant local maxima and group those small-disconnected regions together as one connected region. After Radon Transform, the spatial maxima mapping (SMM) algorithm is designed to match unknown poses and known poses. Finally, the mapping results with the highest score are used to classify the human pose \cite{pose_rt}.

\section{Multimedia in training and education}
Multimedia in training includes a cost-effective system for rehabilitation by utilizing virtual reality. Multimedia in education includes an innovative environment for learners to optimize their learning experience and to deliver portable and scalable education. The multimedia in training and education related papers proposed by MRC include:

\subsection{A Framework for Adaptive Training and Games in Virtual Reality Rehabilitation Environments}
There are many challenges when disabled individuals are using electric power wheelchairs at their training and rehabilitation stages. Especially in children's cases, equipment adjustment and the environment with specialists are essential but limited. Therefore, a new adaptive strategy for training and games in the virtual rehabilitation environment is introduced in this paper. By using virtual reality and a powered wheelchair simulator in training, patients can be more engaged in the training session for a longer period. This proposed rehabilitation system is flexible, low-cost, and focused on indoor environments for safer and effective training. Both interactive and standard training modes are provided to ensure patient engagement. Comparing with other approaches in the area, the clinicians can design, build, and customize the interactive training environments for patients on this system, resulting in a more effective training process with dynamically responding to a user’s action.
Along with the framework based on Bayesian networks, intelligent and automatic adaption to various types and levels of training that patients have at different times is addressed. The same approach can also be applied to other spatial-based training applications. Further extension of remote rehabilitation training capabilities is considered for the future enhancement of the proposed virtual reality simulation \cite{vr_rehabilitation}.

\subsection{Interactive Multimedia for Adaptive Online Education}
In the paper \cite{interactive_multimedia}, I. Cheng et al. proposed an innovative Computer Reinforced Online Multimedia Education (CROME) framework for multimedia education. By integrating the components of learning, teaching, testing, and student modeling, the CROME framework provides the groundwork for more robust and effective architecture to be developed upon. Also, the proposed system's design integrated different aspects of education using 3D items, drag and drop items, logical-mathematical items, language items, and educational games \cite{interactive_multimedia}. Besides, an automatic difficulty level is estimated for selected subjects. Tests are conducted in innovative item formats.

Moreover, questions are created with assistance from computers, and modules are designed for tracking precise concepts. Furthermore, the testing is beyond only subject knowledge. In summary, the CROME design aims to optimize the available multimedia resources and development kits to develop portable and scalable items for education. Also, it ensures that multimedia education would appeal to and entertain learners. The architecture of the system also achieves portability, reusability, scalability, and interoperability.

\section{Conclusion}
The main goal of this survey is to present various multimedia technologies and robust algorithms ranging from daily life to medical research. Realizing the concept of multimedia solutions requires an understanding of various components, e.g. computer vision and data processing. In this survey, we have discussed multimedia technologies with various data processing techniques, adaptive systems, and medical and human-related researches. Through the survey, we have highlighted the benefits and proposed methods presented in the literature from Multimedia Research Centre (MCR). In the long term, our survey is working on promoting collaborations between multiple disciplines.  Along with MCR, we hope to move forwards by exploring more efficient multimedia solutions by combining traditional computer vision methods with deep learning techniques to process and manage multimedia data. 

\bibliographystyle{IEEEtran}
\bibliography{main}
\vspace{-1 cm}
\begin{IEEEbiographynophoto}{Zijian Kuang}
is a machine learning enthusiast who is currently pursuing his MSc. in Computing Science at the University of Alberta. He has 5+ years of work experience in software development and project management with experience working in several applications for government and large organizations.
\end{IEEEbiographynophoto}

\vskip 0pt plus -1fil
\begin{IEEEbiographynophoto}{Xinran Tie}
is currently a student pursuing MSc. in Computing Science with Specialization in Multimedia at the University of Alberta. She is expected to graduate in December 2021, and currently studies on and researches 3 projects from the courses. 
\end{IEEEbiographynophoto}

\end{document}